# Maximizing the Spread of Cascades Using Network Design

**Daniel Sheldon, Bistra Dilkina, Adam N. Elmachtoub, Ryan Finseth, Ashish Sabharwal,**
**Jon Conrad, Carla Gomes, David Shmoys**
Cornell University, Ithaca, NY

**William Allen, Ole Amundsen, William Vaughan**
The Conservation Fund, Arlington, VA

## Abstract

We introduce a new optimization framework to maximize the expected spread of cascades in networks. Our model allows a rich set of actions that directly manipulate cascade dynamics by adding nodes or edges to the network. Our motivating application is one in spatial conservation planning, where a cascade models the dispersal of wild animals through a fragmented landscape. We propose a mixed integer programming (MIP) formulation that combines elements from network design and stochastic optimization. Our approach results in solutions with stochastic optimality guarantees and points to conservation strategies that are fundamentally different from naive approaches.

## 1 INTRODUCTION

Many natural processes — such as the diffusion of information in a social network or the spread of animals through a fragmented landscape — can be described as a network diffusion process or *cascade*. For example, an individual who buys a new product or adopts a new technology may trigger similar behavior in friends; if this process gains momentum it can cascade through a significant portion of a social network. The study of cascading behavior in networks is most familiar in social or epidemiological settings (Goldenberg et al., 2001; Leskovec et al., 2007a; Anderson and May, 1992; Chakrabarti et al., 2008). However, a similar framework called *metapopulation modeling* exists in ecology to describe the occupancy pattern of habitat patches in a fragmented landscape (Hanski, 1999).

One would often like to intervene to steer the course of a cascade toward some goal, e.g., to maximize its spread through the network. However, a cascade is a complex stochastic process and it is typically not possible to directly control the outcome, but only to influence certain underlying or initial conditions, such as where to initiate a cascade. This leaves great uncertainty as to the actual outcome of a given intervention. In this paper, we contribute a new optimization framework to maximize the expected spread of a cascade under a very general class of interventions.

The problem of maximizing the spread of a cascade is of great practical import. In the social network setting, Domingos and Richardson (2001) posed the problem of targeting individuals to maximize the effectiveness of a viral marketing strategy. Kempe, Kleinberg, and Tardos (2003) later showed that for several different cascade models, finding the optimal set of $k$ individuals to initiate a cascade in order to maximize its eventual spread is NP-hard, but because it is a submodular optimization problem, a greedy approach finds approximate solutions with strong performance guarantees.

In a conservation setting, maximizing the spread of a target species through the landscape given a limited management budget is a central problem in the newly emerging field of *computational sustainability* (Gomes, 2009). In this case, the management tools consist of augmenting the network of habitat patches through conservation or land acquisition. Unlike the social network example, the initial locations for the cascade are predetermined by the current spatial distribution of the species, which is difficult to manipulate. Moreover, metapopulation models predict that long-term population dynamics are determined by properties of the landscape much more so than initial occupancy (Ovaskainen and Hanski, 2001).

Motivated by these considerations, we introduce a much more general optimization framework for cascades. In our model, one may choose from a rich set of *management actions* to manipulate a cascade: in addition to choosing where to initiate the cascade, actions may also intervene directly in the network to change cascade dynamics by adding nodes. This model is general enough to also capture adding edges, or increasing the local probability of propagating the cascade.

The objective function is no longer submodular with respect to this more general decision space, so optimization becomes more difficult. We formulate the problem as mixed integer program (MIP) that combines elements from deterministic network design problems and stochastic optimization.

One of our main computational tools is *sample average approximation* (SAA): we find a solution that is optimal in hindsight for a number of *training cascades* that are simulated in advance. The SAA optimum may overfit for a small set of training cascades, but converges to the true optimum with increasing training samples, and provides a stochastic bound on the optimality gap. Moreover, because the set of training cascades are known prior to optimization, SAA allows significant computational savings compared with other simulation-based optimization methods.

In addition to the MIP-based SAA approach, we contribute a set of preprocessing techniques to reduce computation time for SAA and other algorithms that repeatedly reason about a fixed set of training cascades. Our method compresses each training cascade into a smaller cascade that is identical for reasoning about the effects of management actions. Our experiments show that running times are greatly reduced by preprocessing, both for the SAA approach and for two greedy baselines. adapted from Kempe et al. (2003) and Leskovec et al. (2007b). After preprocessing, our SAA problem instances are amenable to solution by branch-and-bound MIP solvers, even though they are instances of an NP-hard network design problem.

We apply our model to a sustainability problem that is part of an ongoing collaboration with The Conservation Fund to optimize the conservation of land to assist in the recovery of the Red-cockaded Woodpecker (RCW), a federally listed rare and endangered species. Unlike heuristic methods that are often used in conservation planning, our method directly models desired conservation outcome. For the RCW problem, we find solutions with stochastic optimality guarantees that demonstrate conservation strategies fundamentally different from those found by naive approaches.

## 2 PROBLEM STATEMENT

We begin by stating the generic optimization problem for *progressive* cascades; these serve as the foundation for all of our modeling. Specifics of our conservation application — which uses a variant called a *non-progressive cascade* — are given in Section 2.3.

A progressive cascade is a stochastic process in a graph $G = (V, E)$ that begins with an initial set of active nodes; these proceed to activate new nodes over time according to local activation rules among neighbors until no more activations are possible. Given a limited budget and a set of target nodes, we wish to select from a set of *management actions* that affect the activation dynamics in order to maximize the expected number of target nodes that become active.

Let $\mathcal{A} = \{1, \ldots, L\}$ be a set of management actions, and let $c_\ell$ be the cost of action $\ell$. Let $\mathbf{y}$ be a *strategy* vector that indicates which actions are to be taken: $\mathbf{y}$ is a 0-1 vector where $y_\ell = 1$ if and only if action $\ell$ is taken. Strategy $\mathbf{y}$ results in a particular cascade process based on the specification of the cascade model and management actions. Let $\{X_v(\mathbf{y})\}_{v \in V}$ be random variables capturing the outcome of a cascade under strategy $\mathbf{y}$ — the variable $X_v(\mathbf{y})$ is 0 or 1 indicating whether or not $v$ is activated. Let $B$ be the budget, and let $\mathcal{T}$ be a set of target nodes. The formal problem statement is

$$\max_{\mathbf{y}} \sum_{v \in \mathcal{T}} E[X_v(\mathbf{y})] \text{ s.t.} \sum_{\ell=1}^{L} c_\ell y_\ell \leq B. \quad (1)$$

In the remainder of this section we discuss the details of the cascade model and management actions.

### 2.1 CASCADE MODEL

Our model is the *independent cascade model* (Goldenberg et al., 2001; Kempe et al., 2003). Consider a graph $G = (V, E)$ with *activation probabilities* $p_{vw}$ for all $(v, w) \in E$. Notationally, we sometimes write $p(v, w)$ for clarity, and adopt the convention that $p_{vw} = 0$ for $(v, w) \notin E$.

A (progressive) cascade proceeds is a sequence of *activations*. To begin, each node in a given source set $\mathcal{S}$ is activated. When any node $v$ is *first* activated, it has a single chance to activate each neighbor $w$. It succeeds with probability $p_{vw}$ independent of the history of the process so far. If the attempt succeeds and $w$ was not already active, then $w$ becomes newly activated, and will have a chance to activate its neighbors in subsequent rounds. If the attempt fails, $v$ will never attempt to activate $w$ again. Pending activation attempts are sequenced arbitrarily.

In their proofs, Kempe et al. (2003) argue that the following procedure is an equivalent, albeit impractical, way of simulating a cascade. For each edge $(v, w) \in E$, flip a coin with probability $p_{vw}$ to decide whether the edge is *live*. Then the set of activated nodes are those that are reachable from $\mathcal{S}$ by live edges. This is tantamount to simulating the cascade, but flipping the coins for each possible activation attempt in advance. Let the subgraph $G'$ consisting of live edges be called the

*cascade graph*. The SAA method presented in Section 4 will optimize over a fixed set of cascade graphs that are simulated in advance.

**Non-progressive cascades**. In a progressive cascade, an activated node remains so forever (even though it only attempts to activate its neighbors once). This is appropriate for social settings where a person adopts a new technology or buys a new product only once. However, patch occupancy in a metapopulation is *non-progressive*: empty habitat patches may become unoccupied and then occupied again many times, and occupied patches may colonize others during any time step, not only upon their first activation. To model this, suppose that all activations are batched in rounds corresponding to discrete time steps, and that an active node $i$ has probability $\beta_i$ of becoming inactive during each time step.

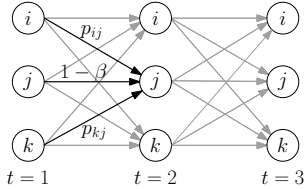

Figure 1: Cascade in layered graph

Kempe et al. showed how to reduce a non-progressive cascade in graph $H = (V, E)$ to a progressive cascade in the *layered graph* $G = (V^T, E')$, where nodes are replicated for each time step, and the nodes at time $t$ connect only to those at time $t+1$ (see Figure 1). Let $v_{i,t} \in V^T$ represent the node $i \in V$ at time $t$. The non-progressive cascade in $H$ is equivalent to a progressive cascade in $G$ with probabilities:

$$p(v_{i,t}, v_{j,t+1}) = p_{ij}, \quad p(v_{i,t}, v_{i,t+1}) = 1 - \beta_i,$$

and $p(v, w) = 0$ for all other $v, w \in V^T$. We think of this as node $i$ at time $t$ activating *itself* at time $t+1$ with probability $1 - \beta_i$, and failing to do so with probability $\beta_i$ so that it becomes inactive. (However it will remain active if it is also activated by a neighbor in the same time step).

## 2.2 MANAGEMENT ACTIONS

Management actions in our model consist of predefined sets of nodes that may be added to the network for a cost. Consider the problem of augmenting a graph $G_0$ on vertex set $V_0$ to optimize for the spread of cascades. Let $V_\ell$ be the set of nodes purchased by action $\ell$, and let $V = V_0 \cup \left(\bigcup_{\ell=1}^L V_\ell\right)$ be the complete set of vertices. Let $G$ be the corresponding graph on vertex set $V$, and assume that activation probabilities $p_{vw}$ for each $v, w \in V$ are known input parameters.

Given a strategy $\mathbf{y}$, a cascade may proceed through nodes that are part of the resulting network, either because they belong to $V_0$ or because some action was taken to purchase the node. Let $V(\mathbf{y}) = V_0 \cup \left(\bigcup_{\ell: y_\ell = 1} V_\ell\right)$ be the set of nodes purchased by $\mathbf{y}$, and let $G(\mathbf{y})$ be the corresponding subgraph of $G$. The cascade process is then fully specified by letting $X_v(\mathbf{y})$ be equal to one if $v$ is reachable by live edges from $\mathcal{S}$ in the subgraph $G(\mathbf{y})$, when each edge $(v, w) \in E$ is chosen to be live independently with probability $p_{vw}$.

A model that purchases sets of nodes is sufficiently general to model other interesting management actions such as the purchase of edges or sources. For example, to purchase edges, modify the graph as follows: replace edge $(v, w)$ by two edges $(v, e)$ and $(e, v)$, where $e$ is a new node representing the edge purchase. Let $p'(v, e) = p(v, w)$, $p'(e, w) = 1$. Then the cascade proceeds from $v$ to $w$ in the new graph with probability $p_{vw}$, only if node $e$ is purchased. Similar ideas can be used to model actions that purchase sources, so that the submodular influence-maximization problem of Kempe et al. can be seen as a special case of our optimization problem.

## 2.3 CONSERVATION APPLICATION

Here we describe how this model of cascades and management actions relates to metapopulations and a conservation planning problem. A metapopulation model is a non-progressive cascade that describes the occupancy of different habitat patches over some time horizon $T$. Node $v_{i,t}$ represents habitat patch $i$ at time $t$. For $i \neq j$, the activation or *colonization probability* $p_{ij}$ represents the probability that an individual from patch $i$ will colonize patch $j$ in one time step. The *extinction* probability $\beta_i$ is the probability that the local population in patch $i$ will go extinct in one time step. We assume that colonization and extinction probabilities are specified in advance, although in practice they are very difficult to know precisely.

Patches are grouped into non-overlapping *parcels* $\mathcal{P}_1, \ldots, \mathcal{P}_L$; some are already conserved, and the others are available for purchase at time 0. For each unconserved parcel $\mathcal{P}_\ell$, there is a corresponding management action with node set $V_\ell$ containing the nodes $v_{i,t}$ for all patches $i \in \mathcal{P}_\ell$ and for all $t$. The set $V_0$ consists of nodes representing patches in parcels that are already under conservation. In other words, the target species may only occupy patches that fall within conserved parcels, and the land manager may choose additional parcels to purchase for conservation. The fact that the species may not exist outside of conserved parcels is a simplification, but there is flexibility in defining "conserved", which is adequate for our purposes.

The sources consist of nodes $v_{i,0}$ such that patch $i$ is occupied at the beginning of the time horizon. The target set is $\mathcal{T} = \{v_{i,T}\}$, i.e., the objective is to maximize the expected number of occupied patches at the end of the time horizon.

### 2.4 NON-SUBMODULARITY

Our problem shares the same objective as the submodular influence-maximization problem. However, with respect to our expanded set of actions, the objective is not submodular. A set function $f$ is submodular if for all $S \subseteq T$ and for all $m$, the following holds:

$$f(S \cup \{m\}) - f(S) \geq f(T \cup \{m\}) - f(T).$$

This is a diminishing returns property: for any $m$, the marginal gain from adding $m$ to a set $T$ is no more than the gain of adding $m$ to a subset $S$ of $T$.

It is easy to see that submodularity does not hold for the expanded set of actions in our problem, and that the greedy solution can perform arbitrarily worse than optimal. This is true because an instance may contain a high payoff action that is only enabled by first taking a low payoff action.

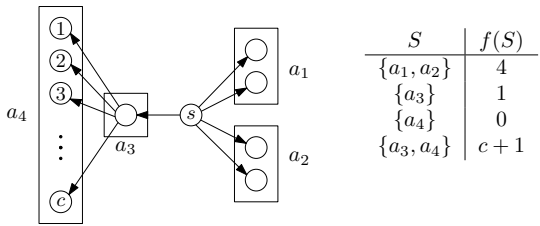

Figure 2: Example of non-submodularity.

Consider the example in Figure 2, with a single source node $s$, unit cost actions, and where activation probabilities are equal to 1 for all edges. Action $a_4$ has payoff of $c$, but only if action $a_3$ is also taken. Hence, the marginal gain of adding $a_4$ to $\{a_3\}$ is greater than that of adding $a_4$ to the empty set, so the objective is not submodular. For a budget of 2, the greedy algorithm will select $\{a_1, a_2\}$ for a total reward of 4, but the optimal solution $\{a_3, a_4\}$ has objective value $c+1$.

## 3 RELATED WORK

**Cascades.** Cascade optimization has also been considered by Leskovec et al. (2007b) and Krause et al. (2008) in their work on optimal sensor placement for outbreak detection in networks. They seek the optimal selection of nodes to *detect* cascades that occur in a network, and show that this problem is also submodular so can be approximated well by a greedy approach. They make improvements to the greedy approach to: (1) provide bounds in the case of non-uniform sensor costs and (2) greatly improve performance by introducing the CELF algorithm, which makes the same selections as a naive greedy algorithm with many fewer function evaluations by taking advantage of submodularity to avoid unnecessary evaluations.

**Stochastic Optimization.** Stochastic optimization problems have long been considered to be significantly harder computational problems than their deterministic counterparts. E.g., it is easy to show even in extremely simple settings that certain classes of stochastic linear programs are #P-hard (Dyer and Stougie, 2006). Only in the last two decades has substantial attention been paid to solving stochastic integer programs as well (see, e.g., the survey of Ahmed (2004)), and the sample average approximation (SAA) has been instrumental in recent advances. The survey of Shapiro (2003) outlines the range of convergence results for SAA that can be proved for a wide swath of stochastic optimization problems. The SAA also yields surprising strong approximation algorithm results, for both structured linear and integer programming problems, as surveyed by Swamy and Shmoys (2006). This approach has recently been applied to other large-scale stochastic combinatorial optimization problems, such as was done in the work of Verweij et al. (2003).

## 4 METHODOLOGY

In this section, we describe our method to solve the cascade optimization problem (1). A major challenge is the fact that the objective itself is very difficult to compute. Even for a fixed strategy $\mathbf{y}$, the problem of computing $E[X_v(\mathbf{y})]$ for a single node $v$ is equivalent to the #P-complete *s-t reliability problem* of computing the probability that two terminals remain connected in a graph with random edge failures (Valiant, 1979).

### 4.1 SAA

The sample average approximation method (Shapiro, 2003; Verweij et al., 2003) is a technique for solving stochastic optimization problems by sampling from the underlying distribution to generate a finite number of scenarios and reducing the stochastic optimization problem to a deterministic analogue. In our setting, instead of maximizing the expected value in Problem (1) directly, the SAA method maximizes the empirical average over a fixed set of samples from the underlying probability space. It is important to note that the random variables $X_v(\mathbf{y})$ can be measured in a fixed probability space defined over subgraphs of $G$ that does not depend on the particular strategy $\mathbf{y}$. In other words, to simulate a cascade in $G(\mathbf{y})$, one can first flip coins for all edges in $G$ to construct a subgraph $G'$ (the cascade graph), and then compute reachability in $G'(\mathbf{y})$.

Let $G'_1, \ldots, G'_N \subseteq G$ be a set of *training cascades* produced in this fashion, and let $\xi_v^k(\mathbf{y})$ be a deterministic value that indicates whether or not node $v$ is reachable in $G'_k(\mathbf{y})$. The sample average approximation of (1) is

$$\max_{\mathbf{y}} \frac{1}{N} \sum_{k=1}^{N} \sum_{v \in \mathcal{T}} \xi_v^k(\mathbf{y}) \text{ s.t.} \sum_{\ell=1}^{L} c_\ell y_\ell \leq B. \quad (2)$$

To encode this as a MIP, we introduce reachability variables $x_v^k$ to represent the values $\xi_v^k(\mathbf{y})$, and a set of linear constraints that enforce consistency among the $\mathbf{x}$ and $\mathbf{y}$ variables such that they have the intended meaning. Let $\mathcal{A}(v)$ be the subset of actions that purchase node $v$. To match our application, we use the following formulation tailored to non-progressive cascades, for which $G'_k$ is guaranteed to be acyclic, though this is not a fundamental limitation.

$$\max_{\mathbf{x},\mathbf{y}} \frac{1}{N} \sum_{k=1}^{N} \sum_{v \in \mathcal{T}} x_v^k$$

$$\text{s.t.} \sum_{\ell=1}^{L} c_\ell y_\ell \leq B$$

$$x_v^k \leq \sum_{\ell \in \mathcal{A}(v)} y_\ell, \qquad \forall v \notin V_0, \forall k \quad (3)$$

$$x_v^k \leq \sum_{(u,v) \in E_k} x_u^k, \qquad \forall v \notin \mathcal{S}, \forall k \quad (4)$$

$$0 \leq x_v^k \leq 1, \ y_\ell \in \{0, 1\}.$$

Consider a fixed $\mathbf{y}$. We say that node $v$ is *purchased* if $\sum_{\ell \in \mathcal{A}(v)} y_\ell \geq 1$. The constraints (3) and (4) together imply that $x_v^k > 0$ only if there is a path in $G'_k$ from the sources $\mathcal{S}$ to $v$ consisting of purchased nodes. In this case, the constraints are redundant and $x_u^k$ may be set to the upper bound of 1 for all nodes $u$ on the path. If there is no path to $v$, then an inductive argument shows that $x_v^k$ must be equal to 0. Otherwise, by (4), there must be some node $u$ such that $x_u^k > 0$ and $(u,v) \in E_k$. By induction, we can build a reverse path from $v$ comprised of nodes $w$ such that $x_w^k > 0$. Such a path must end at a source (recall that $G'_k$ is acyclic), contradicting the fact that $v$ is not reachable.

*Relationship to Network Design.* After sampling the training cascades, our problem is a deterministic *network design problem*: which sets of nodes should be purchased to connect the most targets? Network design is one of the most well-studied classes of deterministic combinatorial optimization problems. Traditionally, these problems arise in applications such as telecommunication networks and supply chains, where there is a given input graph that specifies potential routing links (and nodes) that can be purchased to provide a specified quality of service, and the aim is to do this at minimum cost. Our formulation is similar to standard flow-variable based network design MIPs (e.g., see Magnanti and Wong (1984)), but since the input graphs are acyclic and purchases are made on nodes instead of edges, we can utilize a more compact formulation without any edge variables.

**Algorithm 1**: The SAA procedure.

**Input**: $M$ samples of $N$ training cascades, $N_{\text{valid}}$ validation cascades, $N_{\text{test}}$ testing cascades

**1** Solve $M$ independent SAA problems of size $N$ to produce candidate solutions $\mathbf{y}_1, \ldots, \mathbf{y}_M$, and upper bounds $\bar{Z}_1, \ldots, \bar{Z}_M$. Set $\bar{Z} = \frac{1}{M} \sum_{i=1}^{M} \bar{Z}_i$.

**2** Choose the best solution $\mathbf{y}^*$ from $\mathbf{y}_1, \ldots, \mathbf{y}_M$ by re-estimating the objective value of each using $N_{\text{valid}}$ independent validation cascades.

**3** Compute $\underline{Z}(\mathbf{y}^*)$ using $N_{\text{test}}$ independent testing cascades. The estimated upper bound on the optimality gap is $\bar{Z} - \underline{Z}(\mathbf{y}^*)$.

**Bounding Sub-Optimality of SAA.** The SAA optimum converges to the true optimum of (1) as $N \to \infty$, but for small $N$ the optimal value may be optimistic, and the solution sub-optimal. Verweij et al. describe the following methodology to derive stochastic bounds on the quality of the SAA solution compared with the true optimum (Verweij et al., 2003).

Let OPT be the true optimal value of problem (1), and let $\bar{Z}$ be the optimal value of the SAA problem (2) for a fixed set of $N$ training cascades. For any solution $\mathbf{y}$ (typically the SAA optimum), let $\underline{Z}(\mathbf{y})$ be an estimate of the objective value of solution $\mathbf{y}$ for problem (1) made by simulating $N_{\text{test}}$ cascades in $G(\mathbf{y})$. The bounds are based on the fact that

$$E[\underline{Z}(\mathbf{y})] \leq \text{OPT} \leq E[\bar{Z}].$$

The value $E[\bar{Z} - \underline{Z}(\mathbf{y})]$ is then an upper bound on the optimality gap $\text{OPT} - E[\underline{Z}(\mathbf{y})]$, and the random variable $\bar{Z} - \underline{Z}(\mathbf{y})$ is an unbiased estimate of the upper bound. To reduce the variance, one solves the SAA problem many times with independent samples and lets $\bar{Z}$ be the average of the upper bounds obtained in this way. This also gives many candidate solutions, of which the best is selected using validation samples. The overall procedure is specified in Algorithm 1. Note that if the SAA problem in line 1 is not solved optimally, the upper bound $\bar{Z}_i$ is the best upper found during optimization, *not* the objective value of $\mathbf{y}_i$.

## 4.2 PREPROCESSING

Because the SAA problem optimizes over a fixed set of training cascades that are sampled in advance, it is possible to achieve significant computational savings by preprocessing the cascades to accelerate later computations. For example, we defined the training cascade $G'_k$ to be a subgraph of $G$ obtained by retaining each edge $(u, v)$ independently with probability $p(u, v)$. However, this may be an inefficient representation of a cascade. A more compact representation is attained by simulating the cascade forward from $\mathcal{S}$ so that nodes and edges that are not reachable under *any* strategy are never explored. Such a simulation results in a graph that is equivalent for computing reachability from $\mathcal{S}$ under any strategy $\mathbf{y}$.

In general, during preprocessing of a cascade we will consider arbitrary transformations of the problem data for each training cascade — originally consisting of the tuple $(\mathcal{S}, \mathcal{T}, \mathcal{A}, G'_k)$ — to a more compact representation that preserves computation of the objective for any strategy $\mathbf{y}$. This is done separately for each training cascade resulting in parameters $\mathcal{S}_k$, $\mathcal{T}_k$, and $\mathcal{A}_k$ that are now specific to the $k$th training cascade; the MIP formulation is modified in the obvious way to accommodate this. We first introduce one generalization: let the *reward vector* $\mathbf{r}^k$ replace the target set $\mathcal{T}$ so that the objective is computed as $(1/N) \sum_k \sum_{v \in V^k} r^k_v x^k_v$. Initially, $r^k_v = 1$ for $v \in \mathcal{T}$, and 0 otherwise. There are two elementary preprocessing steps: *pruning* and *collapsing sources*.

*Pruning.* We exclude any nodes that are not reachable from $\mathcal{S}$ by generating the cascade graph via forward simulation from $\mathcal{S}$. We additionally prune all nodes $v$ with no path to $\mathcal{T}$.

*Collapsing Sources.* If there is a path from $\mathcal{S}$ to $v$ consisting exclusively of nodes from $V_0$, then $v$ will be reachable for any strategy $\mathbf{y}$. In this case, $v$ is added to $\mathcal{S}_k$ as a new source. The set $\mathcal{S}$ is collapsed to a single source node $s$ with an outgoing edge $(s, v)$ for each $(u, v) \in E_k$ such that $u \in \mathcal{S}$.

We also contribute a method to find sets of nodes that can be collapsed into singletons because the fates of all nodes in the set are tied together — i.e, for any strategy, whenever one node in the group is reachable, all are. The simplest example of this is a strongly connected component consisting of nodes purchased by the same action(s). In general, let $u \Rightarrow v$ denote the situation in which, for any strategy $\mathbf{y}$, if node $u$ is reachable, then node $v$ is also reachable. There are two basic cases that guarantee that $u \Rightarrow v$:

1. $(u, v) \in E$ and $\mathcal{A}(u) \subseteq \mathcal{A}(v)$. I.e., $u$ links to $v$, and whenever $u$ is purchased $v$ is also purchased.

2. $(v, u) \in E$ and $\nexists w : (w, u) \in E$. I.e., the *only* link to $u$ comes through $v$; hence, any path that reaches $v$ must go through $u$.

The relation $u \Rightarrow v$ is clearly transitive, so to find all pairs such that $u \Leftrightarrow v$, we build a graph with directed edges corresponding to the two basic cases above and compute its strongly connected components (SCCs). Once we have computed the SCCs, we form the quotient graph by collapsing each strongly-connected component $C$ into a single node and updating the edge set accordingly. We must also update the other parameters of the problem. Let $v_C$ be the node corresponding to component $C$. Then: (i) $v_C$ becomes a source if any node in $C$ was originally a source, (ii) the reward of $v_C$ is equal to the sum of rewards for the original nodes in $C$, and (iii) the new action set of $v_C$ is $\mathcal{A}(v_C) = \bigcap_{u \in C} \mathcal{A}(u)$. To justify (iii), note that any strategy under which all of $C$ is reachable must purchase every node in $C$: the set of actions that does this is exactly $\bigcap_{u \in C} \mathcal{A}(u)$.

These preprocessing steps may be repeated until no additional progress is made and we have obtained a reduced cascade. Because preprocessing results in training cascades that are completely equivalent for reasoning about the objective values of strategies, they may be used in conjunction with *any* algorithm.

## 4.3 GREEDY BASELINES

In our experiments, we use two greedy algorithms adapted from Kempe et al. (2003) and Leskovec et al. (2007b) as baselines. Each starts with an empty set of actions, and repeatedly adds the "best" action that does not violate the budget. The algorithm GREEDY-UC (for *uniform cost*) chooses the action that results in the greatest increase in objective value. The algorithm GREEDY-CB (for *cost-benefit*) chooses the action with the highest ratio of increase in objective value to cost.

In each step, the increase in objective value is evaluated for each possible action by simulating $N$ cascades. For our problem instances, it is prohibitively time consuming to simulate new cascades for each objective function evaluation. Instead, we reuse a set of $N$ pre-sampled training cascades as in the SAA method. Simulations in pre-sampled cascades require many fewer edge explorations, because only live edges from the original cascade are considered. We may also apply our preprocessing techniques in this case. The incremental nature of the greedy algorithms allow for an *additional* optimization: after action $\ell$ is selected in some step, every subsequent strategy will include action $\ell$. Hence we may modify the problem by moving the nodes of $V_\ell$ to $V_0$ so that they become part of the "original" graph to be augmented, and then repeat preprocessing. The accumulated speedups from these optimizations are quite significant: as much as 1000x in our experiments.

## 5 EXPERIMENTS

Our application is part of an ongoing collaboration with The Conservation Fund to optimize land conservation to assist the recovery of the Red-cockaded Woodpecker (RCW), a federally listed rare and endangered species. RCW are a "keystone species" in the southeastern US: they excavate tree cavities used by at least 27 other vertebrate species (USFWS, 2003). However, habitat degradation has led to severe population declines, and existing populations are highly fragmented (Conner et al., 2001). As a result, habitat conservation and management are crucial to the continued viability of RCW.

We use the RCW recovery problem as a test bed for the computational approaches developed in this work. The scientific and political issues surrounding endangered species are complex, and great care must be taken when developing models that predict their fate. Although we use real data for this study, some parameter choices and assumptions have not been thoroughly verified with respect to RCW ecology. Hence one should not draw specific conclusions about RCW conservation planning from the particular results presented here. We stress instead the general applicability of the computational model to a variety of specific conservation problems: coupled with diffusion parameters tailored to a given target species, our model can provide decision makers with key information about balancing management options and resource constraints.

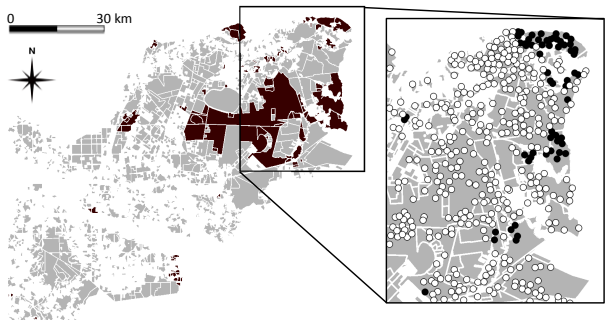

Figure 3: The study area. Left: spatial layout of parcels; dark parcels are conserved. Inset: circles indicate territories; filled circles are occupied. There are no occupied territories outside the inset area.

The study area for these experiments consists of 443 non-overlapping parcels of land within a coastal area in the southeastern United States (see Figure 3); each parcel is at least 125 acres in area (the estimated minimum size to support a RCW territory). The cost to conserve a new parcel is equal to its assessed property value, while already-conserved parcels are free.

RCWs live in small groups in a well-defined *territory* (i.e., patch) that is centered around a cluster of cavity trees where the individual birds roost (Letcher et al., 1998). Presently, the study area contains 63 occupied RCW territories (these determine the sources $\mathcal{S}$), the vast majority of which fall within conserved parcels.

We identify almost 2500 *potential* RCW territories satisfying minimum habitat and area requirements, using a 30 by 30 km habitat suitability raster of the study area that is calculated using land cover type, hurricane/climate change risk, and development risk. The fact that these territories are specified in advance and assumed to exist at time zero is a simplification, but compatible with RCW ecology and management strategies. Because of restrictive requirements for cavity trees (live old-growth pines 80 to 150 years old) and significant time investments to excavate cavities (one to six years), the locations of territories are stable over time: it is far more common for an individual to fill a vacancy in an existing territory than to build a new one (Letcher et al., 1998). Further, it is a common management practice for land managers to drill or install artificial cavities to construct a new territory (USFWS, 2003), which is compatible with the assumption that potential territory locations are chosen in advance of them becoming populated.

The parcel composition can be seen on the left of Figure 3, while the location of active and potential RCW territories can be seen on the right of Figure 3. Given a specific budget, the goal is to decide which parcels to conserve in order to maximize the expected number of active territories at a 100 year time horizon.

To model the dispersal process among RCW territories, we utilize a simple parametric form for colonization probabilities based on previous theoretical work about metapopulations, and the parameter values are adapted to loosely match an individual-based model for the RCW (Letcher et al., 1998). In particular, the probability that some individual from an active territory $i$ colonizes an unoccupied territory $j$ in one year is computed according to Equation 5.

$$p(i,j) = \begin{cases} 1/C_i & d(i,j) \leq r_0 \\ \alpha \exp(-\gamma \cdot d(i,j)) & d(i,j) > r_0 \end{cases} \quad (5)$$

When the distance $d(i,j)$ between territories $i$ and $j$ is within the species foraging radius $r_0$, the colonization probability is inversely proportional to the number $C_i$ of neighboring territories within distance $r_0$. When $d(i,j) > r_0$, the colonization probability decays ex-ponentially with distance. In our study, the foraging radius is $r_0 = 3$ km, and the other parameter values are $\alpha = 0.1$ and $\gamma = 7.69\text{e-}4$. The single-year extinction probability is $\beta_i = 0.29$ for all territories $i$.

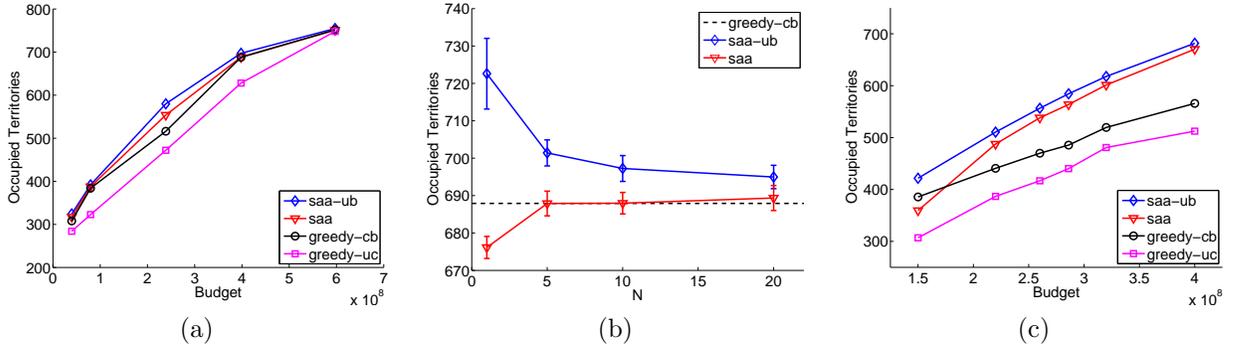

Figure 4: (a) Objective values for SAA and GREEDY, and SAA upper bound, for various budgets. (b) Upper and lower bounds on objective as a function of training size N. (c) Results for MOD instance, K=20.

## 5.1 PERFORMANCE AND BOUNDS

We compared the quality of the solutions that the algorithms SAA, GREEDY-UC and GREEDY-CB found for our test instance for a range of different budgets. For the algorithm SAA, we solved $M = 50$ SAA problems on samples of size $N = 10$ training cascades using the CPLEX optimization software, with the best solution chosen using $N_{\text{valid}} = 500$ validation cascades, and finally evaluated using $N_{\text{test}} = 500$ test cascades (see Algorithm 1). The greedy algorithms were run with 100 training cascades and then evaluated using the same 500 test cascades used for SAA. Performance is measured as the number of occupied territories at the end of 100 years, averaged over the test cascades.

In Figure 4, we plot performance versus budget for the three algorithms, and also the stochastic upper bound ('saa-ub') obtained from SAA. The SAA solutions are very close to this upper bound which indicates that they are essentially optimal. The SAA algorithm outperforms both greedy approaches, although GREEDY-CB also performs very well in this case. Later we will see that certain problem instances exhibit a much bigger performance gap between SAA and the greedy algorithms. Also note that GREEDY-CB outperforms GREEDY-UC, which indicates the importance of considering the benefit of each parcel in conjunction with its cost.

To test whether the number of training cascades we used was sufficient, we evaluated the performance as a function of $N$ at a fixed budget of \$398M (10% of the total parcel cost). Figure 4(b) shows the estimated upper and lower bounds on the objective obtained for each training size. Note that the gaps are quite small:

for $N = 10$ the upper bound is 696.07 and the lower bound is 687.97, only a 1.16% gap. Moreover, the gap does not decrease significantly for $N > 10$, indicating that $N = 10$ is a large enough sample size for SAA to obtain high quality solutions. The error bars in the figure represent 95% confidence intervals that are computed over the $M$ training samples averaged to obtain the upper bound, and the $N_{\text{test}}$ cascades for the lower bound. These indicate high confidence that SAA is close to the *true* optimum for the stochastic optimization problem. Error bars for Figure 4(a) are of similar size, but are omitted because they are too small to be seen relative to the scale of the figure.

## 5.2 PREPROCESSING

Preprocessing is critical to the running time of our algorithms. Our average training cascade has 44K nodes *after* pruning (of 256K possible nodes). The additional preprocessing steps reduce the average cascade size to under 20K nodes and 43K edges. We measured the running time savings by solving SAA instances ($T = 20$, $N = 10$) with and without preprocessing (see Figure 5(a)). Running time is typically reduced by a factor of 3–10 for the range of budgets.

We also evaluated the benefits of preprocessing on the running time of the greedy algorithms, as illustrated in Figure 5(b). The lines are labeled as follows: 'fresh' is the variant of the greedy algorithm that evaluates the marginal benefit of each action by simulating a fresh set of training cascades each time; 'reuse' is the variant where training cascades are simulated in advance as in SAA; 'reuse+pre' also includes preprocessing of the training cascades; and 'reuse+pre+repeat' reapplies the preprocessing step each time the greedy algorithm commits to a management action. The results indicate the dramatic runtime savings accrued by: (1) generating training cascades in advance and (2) preprocessing cascades to reduce their size.

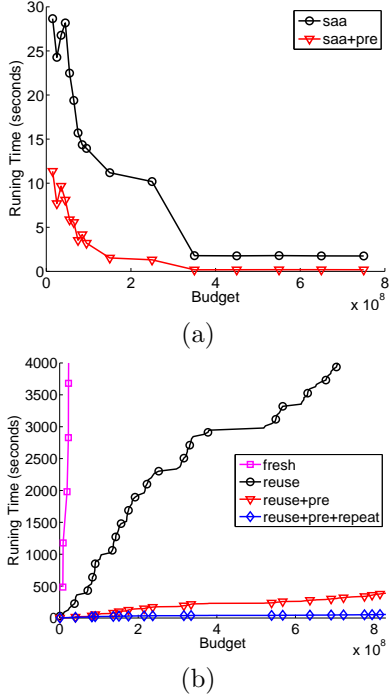

Figure 5: Run times with different levels of preprocessing: (a) SAA, (b) GREEDY-UC.

## 6 DISCUSSION

In Section 2.4, we presented an example that showed that the greedy algorithm can perform arbitrarily worse than optimal. However, in Figure 4(a), GREEDY-CB is often competitive with SAA. Upon investigation, we found that our study area is very well suited to a myopic strategy such as greedy. The diffusion behavior of the RCW given by Equation (5) is such that short range colonizations with hops smaller than the foraging radius $r_0$ are much more common than long range colonizations. Hence, any good solution is effectively constrained to buy parcels that are contiguous with the initially occupied territories, so new parcels can be colonized by short hops. However, the greedy algorithms build outwards myopically, never caring what lies beyond the current parcel they are purchasing, while the SAA algorithm is capable of setting goals, e.g., to build a path from the sources to another area that is highly favorable. However, one can see from Figure 3 that the most favorable area is actually very close to the sources: there is a large block of conserved, but unoccupied, territories in the northeast quadrant of the study area that can support an increased population at no additional cost if they are connected to the sources. Hence, due to the proximity, the greedy algorithm easily discovers a near-optimal solution even though it is behaving myopically.

It is easy to imagine real conservation scenarios where this is not the case. We have modified the problem instance to demonstrate this point, by taking many of the parcels in the northeast quadrant out of conservation and assigning them costs, and then making many of the parcels farther to the west and southwest conserved (i.e., available at no cost). This reflects a situation where an existing population is located at some distance from a reservoir of conservation land that is suitable for habitation, or could be made suitable by low-cost management strategies. All other details of the problem such as colonization probabilities and time horizon remain the same.

In Figure 4(c), we evaluate performance of SAA, GREEDY-CB and GREEDY-UB across different budgets for the modified instance. In this case, unlike in Figure 4(a), there is a substantial performance gap between SAA and both greedy approaches.

Figure 6 illustrates the actual strategies of GREEDY-CB and SAA on this modified instance. We can see qualitatively different strategies between the two. In this case, GREEDY-CB continues to follow the myopic conservation strategy of building outward from the initial population, while SAA recognizes that there is a high available payoff by connecting to existing conservation land and builds a path in that direction.

## 7 CONCLUSION

In this work, we addressed the problem of maximizing the spread of cascades under budget restrictions. Our problem formulation allows a powerful class of management actions that add nodes to the network. However, the cascade is a complex stochastic process so the outcome of any particular action is uncertain, and unlike other cascade optimization problems, this one cannot be provably approximated by a naive greedy approach. We proposed a sample average approximation approach to reduce the stochastic problem to a deterministic network design problem, while still retaining stochastic optimality guarantees. We evaluated our methodology on a key problem from the field of computational sustainability: spatial conservation planning for species population growth. The SAA approach scaled well to this large real-world instance and found better solutions than greedy baselines while also providing optimality bounds. Moreover, preprocessing techniques resulted in a dramatic runtime speedup for both the SAA and greedy algorithms. Most importantly, our results show that the optimal solutions generated by the SAA approach can be qualitatively different than the ones obtained by a myopic greedy approach. A promising avenue of future research is to evaluate this methodology for other conservation and cascade optimization problems.

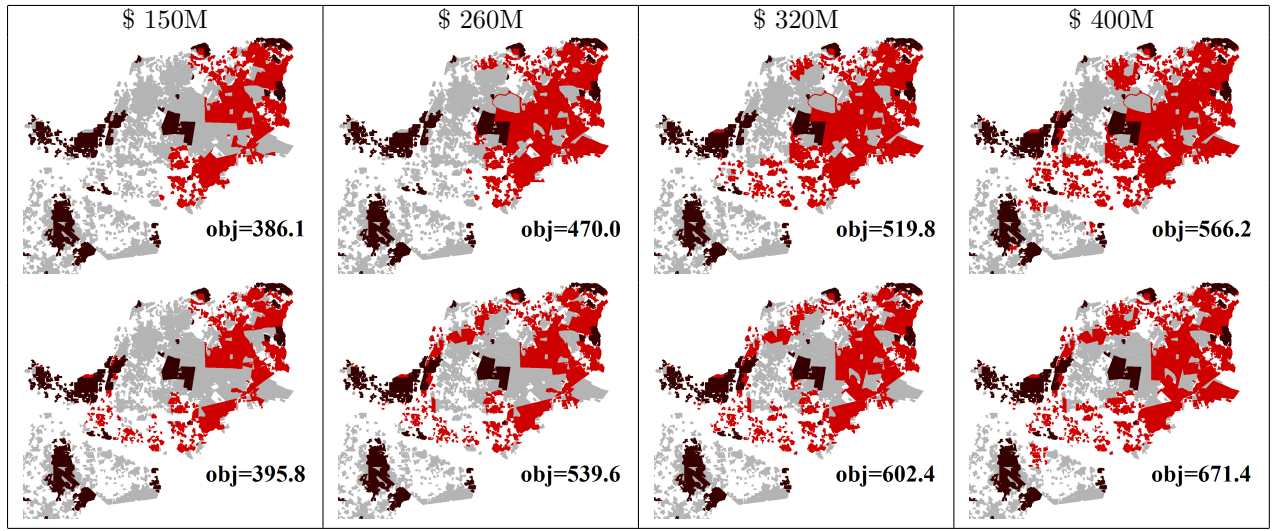

Figure 6: Illustration of conservation strategies for different budget levels obtained by GREEDY (top row) and SAA (bottom row). The expected number of active territories is given for each solution.

### Acknowledgments

This work was supported by the National Science Foundation (grants IIS-0832782, IIS-0514429, and DBI-0905885) and the Air Force Office of Scientific Research (grant FA9550-04-1-0151).